# Cellular Automata Simulation of Grain Growth of Powder Metallurgy Nickel-Based Superalloy

Shasha Liu[a,b], Yiling Jiang[a,b], Ronggui Lu[a,b], Xu Cheng*[a,b], Jia Li[a,b], Yang Chen[c], Gaofeng Tian[c]

a. School of Materials Science and Engineering, Beihang University, 37 Xueyuan Road, Beijing, 100191, China

b. National Engineering Laboratory of Additive Manufacturing for Large Metallic Components, 37 Xueyuan Road, Beijing, 100191, China

c. Beijing Institute of Aeronautical Materials, Beijing, 100095, China

**Abstract**: Primary γ′ phase instead of carbides and borides plays an important role in suppressing grain growth during solution at 1433K of FGH98 nickel-based polycrystalline alloys. Results illustrate that as-fabricated FGH98 has equiaxed grain structure and after heat treatment, grains remain equiaxed but grow larger. In order to clarify the effects of the size and volume fraction of the primary γ′ phase on the grain growth during heat treatment, this paper establish a 2D Cellular Automata (CA) model based on the thermal activation and the lowest energy principle. The CA results are compared with the experimental results and show a good fit with an error less than 10%. Grain growth kinetics are depicted and simulations in real time for various sizes and volume fractions of primary γ′ particles work out well with the Zener relation. The coefficient $n$ value in Zener relation is theoretically calculated and its minimum value is 0.23 when the radius of primary γ' is 2.8μm.

Keywords: grain growth, cellular automata, nickel-based superalloy, microstructure

**Introduction**

Powder metallurgy (PM) nickel-based superalloys have excellent properties such as high temperature resistance, corrosion resistance and oxidation resistance, etc. They are widely used in structural components under high-temperature, such as aero-engine turbine disks[1-3]. Powder metallurgy nickel-based alloys are generally composed of FCC γ matrix, ordered structure $L1_2$ γ' phase ($Ni_3Al$) and a small amount of carbides and borides[4]. The size, morphology and distribution of γ′ phase are related to the working conditions during hot isostatic pressing and heat treatment process[5], which will affect the grain structures. Normally, the second phase (larger precipitations) in the alloy can effectively delay the migration of grain boundaries and thus decelerate the grain coarsening[6, 7].

Due to the difficulty in the in-situ observation of grain growth during heat treatment, microstructure evolution of the alloy is normally analyzed by the means of simulation[8], including the Monte Carlo method[9], phase field method[10], vertex method[11] and cellular automata (CA) method[12]. Among these various calculation methods, the cellular automata (CA) method has been mostly widely used, which includes grain growth model[13, 14], recrystallization[15, 16], phase transition[17]. However, the CA simulation of grain growth is rarely studied in the presence of second phase particles. Raabe et.al[18] studied the recrystallization texture of steel plates using the CA simulation method, and they discussed the effect of the volume fraction of second phase on the microstructure and recrystallization texture of steel. Han et.al[19] investigated the effects of different initial grain size, volume fraction and radius of the second phase particles on the microstructure of low carbon steel through the CA model based on grain boundary curvature, and introduced the initial grain size to predict the limit grain size based on the Zener formula[20].

For polycrystalline materials with high volume fraction of second phase particles, the strong pinning effect of these particles can be observed at the grain boundaries. In fact, the presence of second phase plays a decisive role in controlling the grain size, which contributes to the improvement of the mechanical properties, e.g. the yield strength of the alloy is higher with smaller grain size. Taking PM nickel-based superalloy (FGH98) as research object, this paper combines experiments and simulations to study grain growth in the presence of primary $\gamma'$ phase. Simulations are performed with various radius and volume fractions of primary $\gamma'$ phase, and the influence of which on microstructure evolution and final grain size are discussed subsequently.

**1 Experimental materials and methods**

The nickel-based superalloy (FGH98) was first powdered by the plasma rotating electrode (PREP) method. Afterwards, plate-like sample with a diameter of 100 mm was obtained by hot isostatic pressing (HIP) and isothermal forging processes. Five 10mm×10mm×10mm samples were extracted from the as-forged plate and then solution treated at 1433K (the dissolution temperature of the primary γ′ phase[21]) for different times of 30min, 45min, 60min, 75min and 100 min.

After preparation, the samples were mechanically grinded, polished and etched with a chemical solution of 5gCuCl$_2$+100mlHCl+100mlC$_2$H$_5$OH, and then examined by Leica DM4000 optical microscope (OM). The samples were then re-prepared by electro-polishing with 20%H$_2$SO$_4$+80%CH$_3$OH (voltage 25~30V, time 15~20s) and electro-etching with 170 mlH$_3$PO$_4$ +10 mlH$_2$SO$_4$+15gCrO$_3$(voltage 2~5V, time 2~5s) for the purpose of observing γ′ morphology using Hitachi S4800 field emission scanning electron microscope. Image-Pro Plus 6.0 was applied to measure the radius and volume fraction of the primary γ' phase, carbides and borides.

## 2 Cellular Automata Model

### 2.1 Simulation conditions

In this CA model, 300×300 square grid was used, corresponding to an actual space of 120μm×120μm. The cell in the simulation was randomly assigned a number to represent its orientation and cells with the same orientation belong to the same grain. Grain boundaries were determined according to adjacent grains with different orientations. In order to avoid grain coalescence, each grain in the CA model was endued a different direction which will not change unless the grains shrink and disappear during grain growth. The primary γ′ phases with different radius and volume fractions are distributed randomly in the matrix. In order to distinguish the primary γ′ from the matrix, the orientation of the former was assigned a fixed value in this simulation. It is worth noting that we assumed that the primary γ′ phase will neither grow up nor dissolve during the simulation process.

### 2.2 Model transformation rules

The rate of grain-boundary migration has a great influence on grain growth. The rate of grain growth is related to migration rate and driving force, and the rate of boundary migration is related to temperature. The expressions are as follows[14]:

$$v = mF$$

$$m \propto -\frac{Q}{RT}$$

where $v$ and $m$ are the velocity and rate of grain-boundary migration, respectively. $F$ is the driving force, $Q$ is boundary migration energy and $R$ is the molar gas constant. The migration rate will increase as heat treatment temperature increases. Revising on the basis of the thermal activation mechanism of austenite growth CA model[14], the formula of core cell transformation probability is as follows:

$$P_1 = c \cdot exp - \frac{Q}{RT}$$

where $c$ is a constant, and we assume that the boundary migration energy is a fixed value.

The driving force of grain growth is the reduction of grain boundary energy. The transition probability of the core cell is calculated using the lowest energy principle. Assuming interfacial energy is isotropic, interfacial energy of core cell i can be calculated by Hamiltonian function as follows[14]:

$$E_i = J \sum_{k}^{n} [(1 - \delta_{C_i} \delta_{C_k})(1 - f(k))]$$

where $J$ is a measurement of grain boundary energy, which is taken as 1 in this paper. $n$ is total number of the neighbors of the core cell, and its value is 8. $k$ is the kth neighbor of core cell $i$. $\delta$ is the Kronecher symbol, and $C_i$ and $C_k$ are the orientations of cells $i$ and $k$, respectively. $f(k)$ is a function related to the second phase particle. When the kth cell is a second phase particle pinning the grain boundary, the value of $f(k)$ equals to 1, as shown in Figure 1(a). When the kth cell is not a second phase particle or does not pin the grain boundary, the value of $f(k)$ is 0, as shown in Figure 1(b).

After the orientation of the core cell $i$ randomly translates to an arbitrary neighbor cell $j$ among its eight neighbors (C1, C2, C3, C4, C6, C7, C8 and C9, seen in Figure 2), the transformed grain-boundary energy is recorded as $\Delta E$:

$$\Delta E = E_j - E_i$$

According to the lowest energy principle, the transition probability $P_2$ of cell $i$ is calculated based on the following formula:

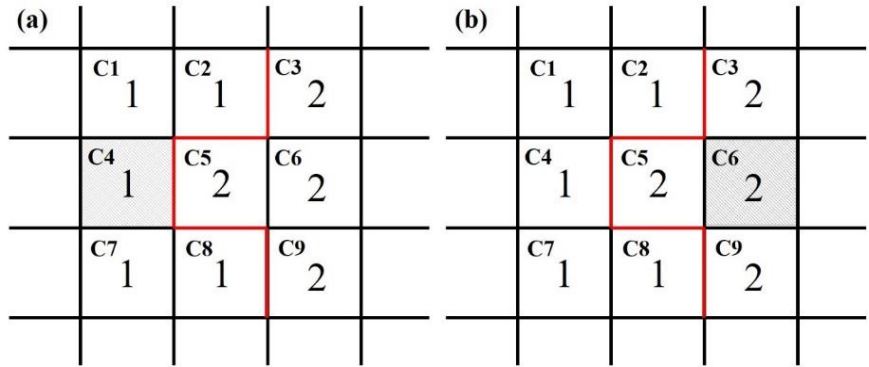

$$P_2 = \begin{cases} 0 & \Delta E \geq 0 \\ 1 & \Delta E < 0 \end{cases}$$

Figure 1 Schematic diagram of the second phase distribution: (a) second phase particles pin the grain boundary; (b) second phase particles are in the grain.

## 3 Results and discussion

### 3.1 Microstructure

The EBSD results in Figure 2 illustrates that as-fabricated FGH98 alloy has equiaxed grain morphology with random grain orientation. The SEM image of the microstructure of FGH98 alloy in Figure 3 indicates that after heat treatment, the morphology of the alloy remains equiaxed with few change observed. In addition, it is obvious that the microstructure is composed of γ matrix, primary γ' phase, carbides and borides[8]. Figure 3(a) and (b) shows that the primary γ′ phase, with irregular morphology, is mostly distributed at grain boundaries and plays a key part in hindering the grain growth [8]. Song et.al[22] found that decrease in the volume fraction of primary γ' phase could lead to the reduction of curvature of grain boundaries and eventually cause the grain coarsening. According to statistics in Figure 1(c) and (d), the radii of primary γ' and the carbides and borides are ranging from 0.5 to 3μm and 0.04 to 0.48μm, respectively. Given that the carbides and borides, more than half of which have a size less than 0.25 μm, are around six times smaller than primary γ′ and mostly appear in grain interiors. It can be assumed that carbides and borides scarcely have pinning effects on the movement of boundaries during heat treatment, and the grain growth behavior during heat treatment are mainly affected by the primary γ′ phase[8].

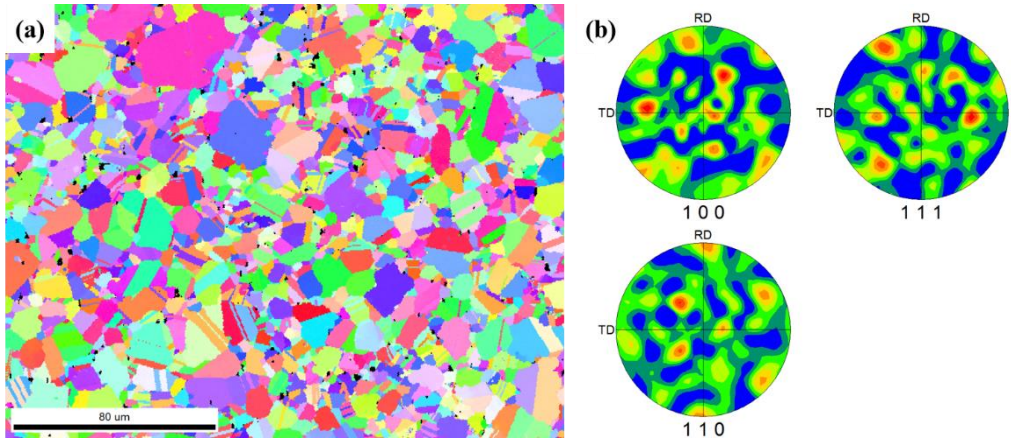

Figure 2 EBSD results of as-fabricated FGH98.

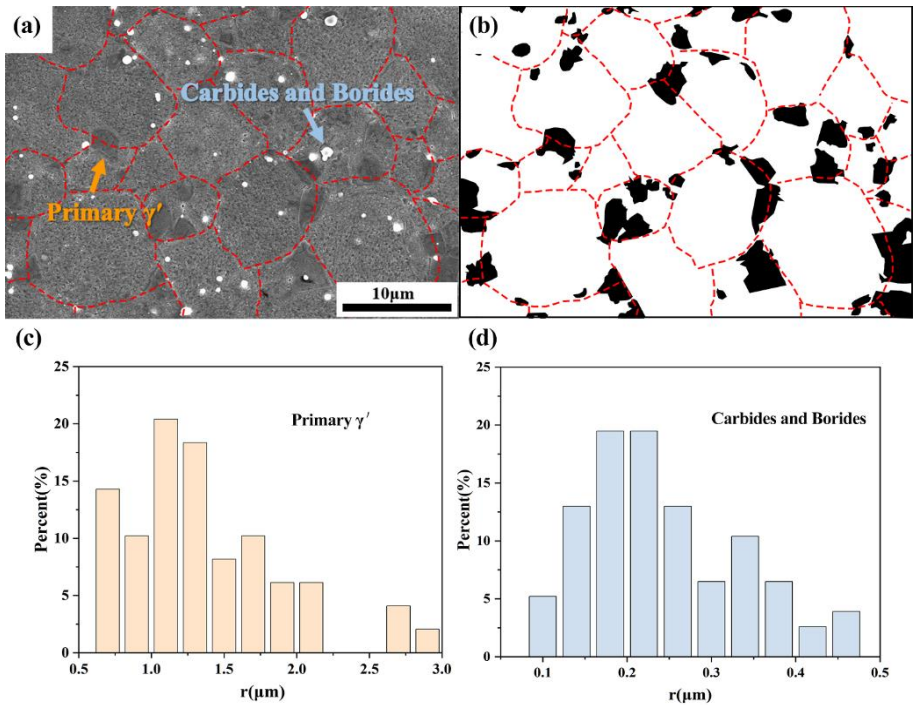

Figure 3 (a) SEM image of solution treated FGH98; (b) the morphology of primary γ′ phase; (c) radius distribution of primary γ′ phase; (d) radius distribution of the carbides and borides.

The radius and volume fraction of the primary γ′ phase of the samples within different holding times were measured and the results are shown in Figure 4. As the holding time increases, both the radius and volume fraction of the primary γ' of samples decrease. At the holding time of 30min, the average radius of primary γ′ is 1.4μm which decreases by about 40% to 0.85μm when increasing the holding time to 100min. Similarly, the volume fraction of primary γ′ drops substantially by 79% over the period, falling from 15% to 3%. Therefore, it can be concluded that the primary γ' phase gradually dissolves when extending holding time.

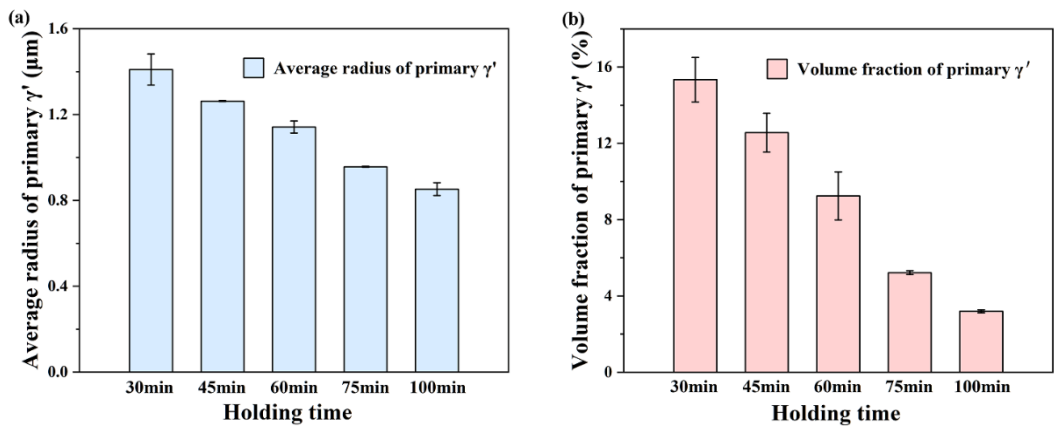

Figure 4 (a) The radius and (b) volume fraction of the primary γ' phase at different holding times.

The top-right insets in Figure 5(a) to (e) are microstructures of FGH98 at different solution time. The grain growth during heat treatment is because that as holding time extends, radius and volume fraction of primary γ' decrease which means primary γ' partially dissolves, failing to pin the grain boundaries effectively. Noticeably, it can be clearly seen that the average grain size slowly increases from 7.6μm at 30min to 11.2μm at 60min, see Figure 5(f). However, when the holding time reaches to 75 min, the average size increases remarkably from 11.2μm at 60min to 22.6μm, almost doubling, which can reach to about 31μm when the holding time extends to 100min. According to Figure 4, the radius of primary γ' at 60min and 75min is 1.1μm and 1.0μm respectively, decreasing by around 9%, whereas the volume fraction of primary γ' is 9.2% and 5.2%, decreasing by about 44%. Therefore, the acceleration in grain growth since 60min can be mainly owing to the losing of pining effect of primary γ' caused by the decrease of volume fraction and the quantitative relation is exemplified later.

In terms of size distribution, the grain size of 30min-treated (Figure 5(a)) and 45min-treated sample (Figure 5(b)) are all below 20μm, distributed relatively uniformly, and for both samples, grain size in the range of 6 and 7.5μm accounts for the largest proportion, which is around 31% and 25% for 30min and 45min, respectively. Compared with the former two treatment states, the particle size distribution in the other three is more concentrated and the maximum proportions are all around 35%. However, the average size of grains with the highest percentage increases as expected from 9.8μm for 60min-treated (Figure 5(c)) to 14.5μm for 70min-treated (Figure 5(d)) and peaks at

21.4μm for 100min-treated sample (Figure 5(e)).

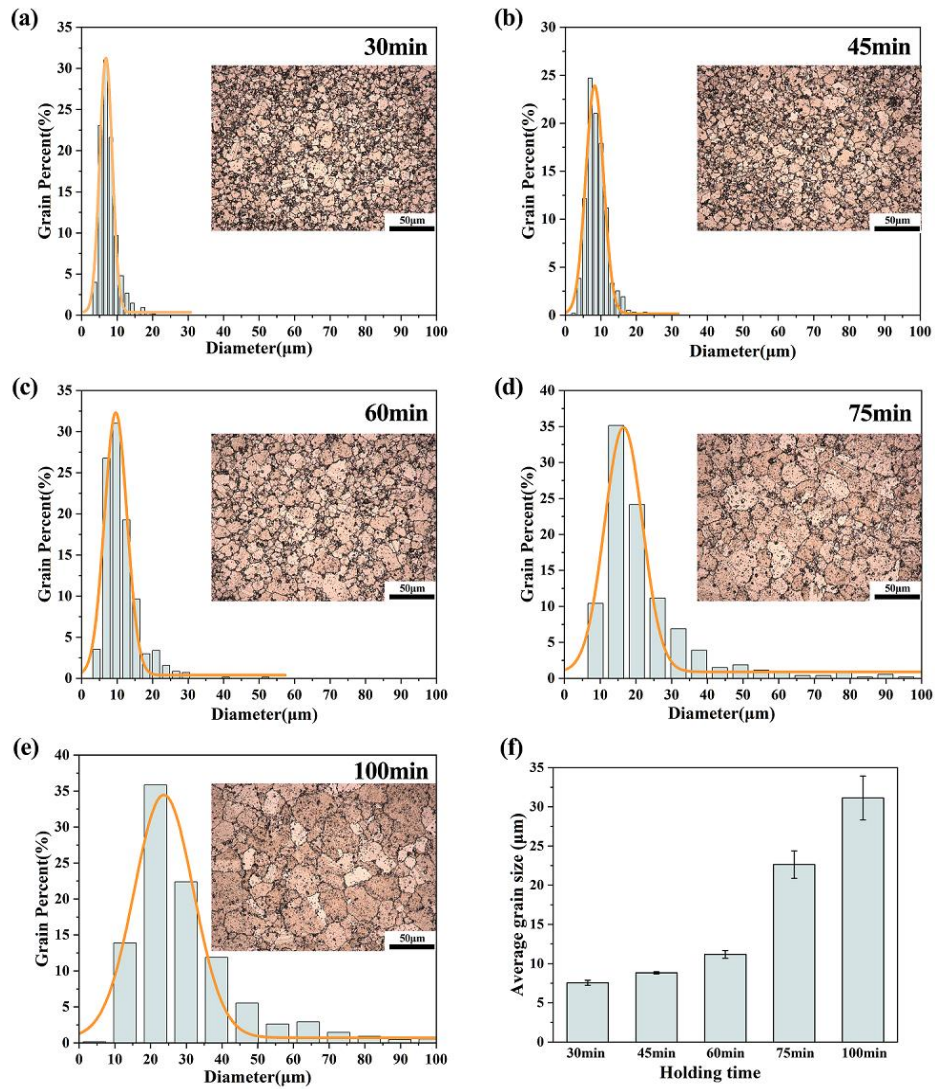

Figure 5 Grain size distribution histograms of FGH98 superalloys with microstructures at image top right at different holding times of (a) 30min, (b) 45min, (c) 60min, (d) 75min and (e) 100min; (f) average grain size histogram.

### 3.2 Cellular Automata Simulation

3.2.1 Simulation model validation

As discussed above, due to the relatively small size of carbides and borides, the paper only takes into account the effects of the primary γ' phase on grain growth. And in order to simplify the model, the shape of it is considered as spherical, with an average size referring to the experimental data.

Figure 6 compares the simulated and experimental microstructures under different holding times. The radius and volume fraction of primary γ' in the simulation are set according to that in the experiment in Figure 4. In Figure 6(a) to (e), the alloy has an equiaxed grain structure, and there is no abnormal grain growth, similar to previous research[18, 23]. Figure 6(f) gives an unambiguous comparison of the average grain size between experiment and simulation. These two results are essentially identical and the maximum error between them is only 8.7% for 60min holding time.

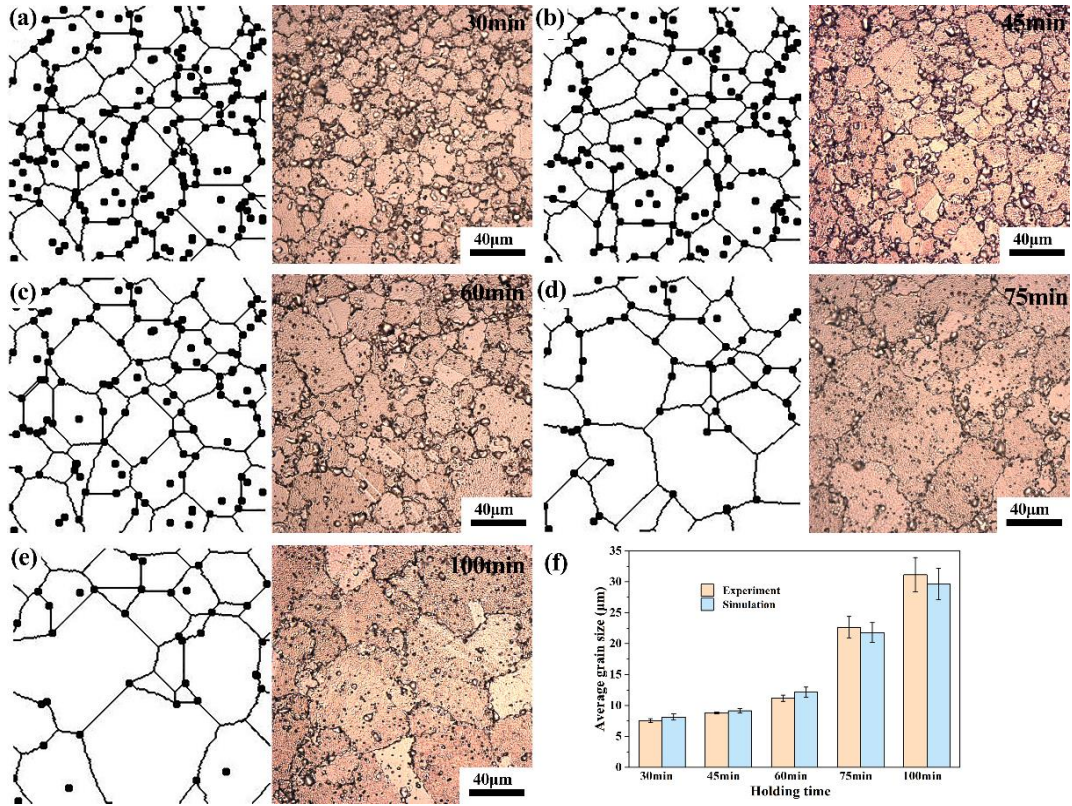

Figure 6 Comparison of simulated and experimental microstructures of FGH98 superalloys for different holding times of (a) 30min, (b) 45min, (c) 60min, (d) 75min and (e) 100min; (f) comparison histogram of average grain size.

Figure 7 depicts the microstructure evolution of nickel-based superalloys with primary γ' phase at a fixed radius (1.2μm) but various volume fractions at different CAS. The increase in CAS can be seen to result in grain growth, whereas the growth rate slows down while increasing the volume fraction of primary γ′ phase. When the fraction reaches to 10%, few changes in grain size is observed (Figure 7(d)). Hence, a large volume fraction of second phase particles will lead to an increasing difficulty of grain boundary moving[24].

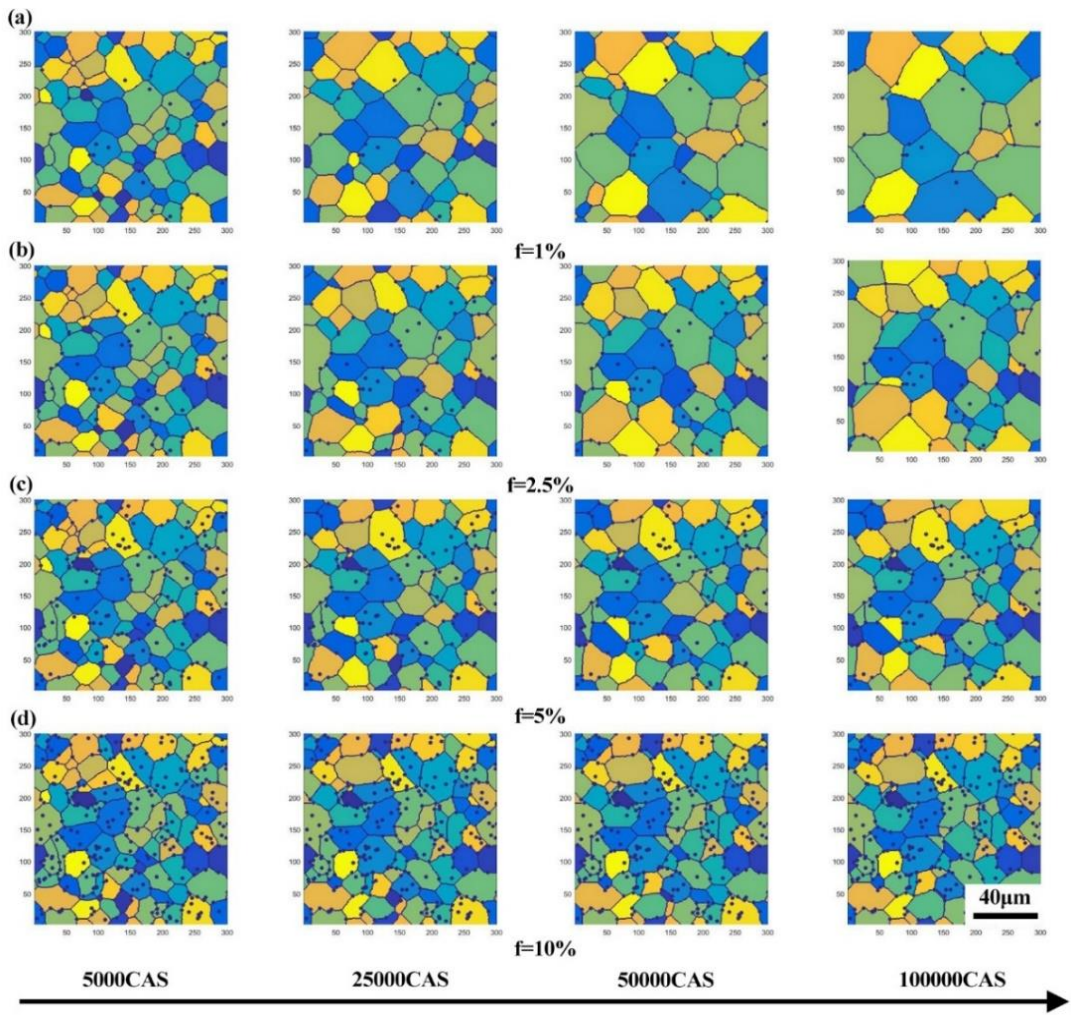

Figure 7 Microstructure evolution of FGH98 superalloys with primary γ' phase at a fixed radius (1.2μm) but different volume fractions at different CAS: (a) f=1%, (b) f=2.5%, (c) f=5%, and (d) f=10%.

Figure 8 shows the microstructure evolution of nickel-based superalloys with primary γ' phase at a fixed volume fraction (5%) but different radii at different CAS. As it can be seen, grain size of the alloy increases with the increase of the CAS. When the radius of primary γ' phase is 1.2μm (Figure 8(a)), the number of grains is almost identical at CAS of 50000 and 100000, indicating an imperceptible grain growth. However, following the addition of the radius to 2.8μm (Figure 8(c)), the grain grows the fastest among three and the number of grains at 100000 CAS is only one third of that at 50000 CAS. Therefore, the simulation results indicated that small second phase particles are more efficient in pinning grain boundaries than large ones with a certain volume fraction.

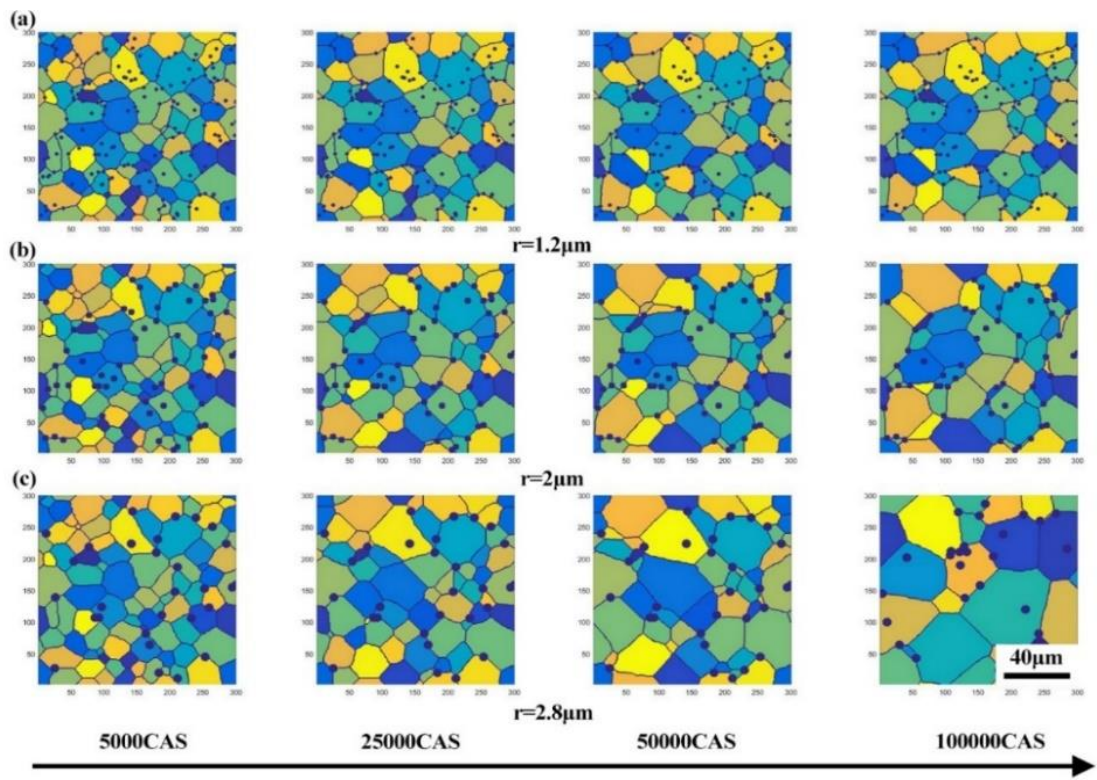

Figure 8 Microstructure evolution of FGH98 superalloys with primary γ' phase at a fixed volume fraction(f=5%) but different radii at different CAS: (a) r=1.2μm, (b) r=2μm and (c) r=2.8μm.

3.2.2 Grain growth kinetics

Figure 9 depicts the grain growth kinetics with fixed particle radius (r=1.2μm) but graded volume fractions of primary γ' phase. Within first 50000CAS, there is apparent growth of the average grain size seen in each volume fraction, and then the growth becomes slower gradually. Especially with the increase of volume fraction, this trend is more pronounced, which means that larger volume fraction of primary γ' exerts their inhibition effect on grain growth in a more efficient way.

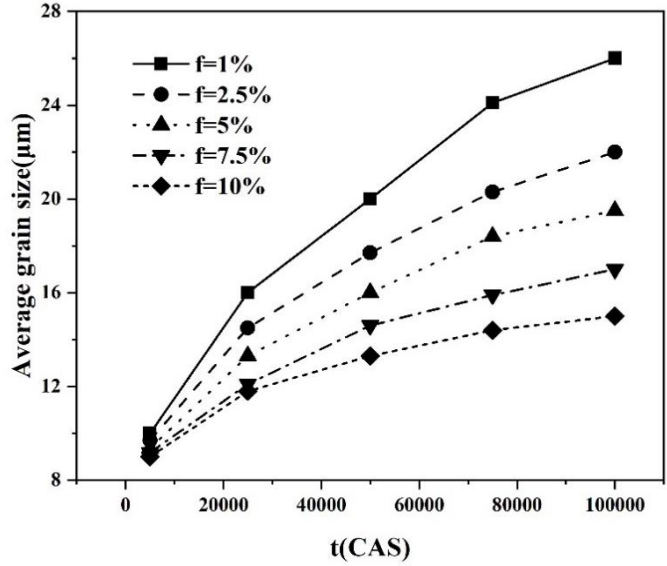

Figure 9 Grain growth kinetics with fixed particle radius (r=1.2μm) but graded volume fractions of primary γ' phase.

The relationship between the grain size, the volume fraction and radius of the second phase has been studied extensively and the general form is the Zener equation[25] as follows:

$$\frac{d_{lim}}{r} = \frac{k}{f^n}$$

where $r$ and $f$ are the average radius and volume fraction of primary γ' particles, respectively. $k$ and $n$ are constants, the values of which vary in different studies[26, 27], depending on assumptions based on shapes and properties of second phase particles. In the paper, we take the limiting grain size, $d_{lim}$, as average grain size after heat treatment. As shown in Figure 10, it turns out that the relationship in FGH98 superalloys fits well with the Zener equation, which proves that primary γ' particles play the main role of pinning grain boundaries and that neglecting carbides and borides has not much effects on simulated results. The specific values of $k$ and $n$ are obtained by nonlinear fitting, which varies for different radii of the primary γ' particles. To be more specific, the value of $n$ decreases as the radius of the primary γ' increases. When the radius reaches to 2.8μm, the value of $n$ is minimum, 0.23. Under the same volume fraction, the number of the primary γ' increases as the radius of the primary γ' decreases, enhancing the grain boundary pinning effect and impeding grain growth.

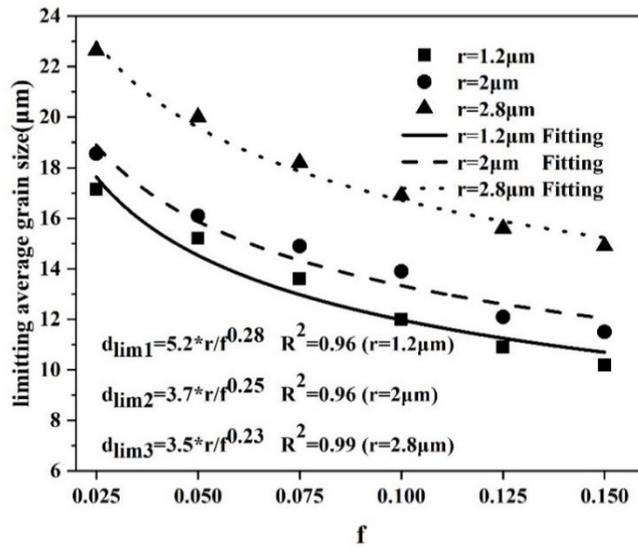

Figure 10 The fitting curve of volume-fraction dependence of limiting average grain size with different radii of primary γ' particles.

**4 Conclusion**

(1) FGH98 nickel-based polycrystalline alloys have equiaxed grains both as-fabricated and solution-treated. During the heat treatment of FGH98, the primary γ' phase instead of carbides and borides can effectively hinder grain growth. With the increase of holding time, the radius and volume fraction of the primary γ' phase gradually decrease, and the grain size experiences an associated increase. The growth kinetics of grains with primary γ' is discussed by 2D cellular automaton based on the mechanism of thermal activation and the lowest energy. The error between the CA (2D Cellular Automata) and experimental results is less than 10%, which indicates the simulation can accurately predict grain growth behavior.

(2) When the radius of primary γ' particles is fixed and the volume fraction is higher than 5%, the increasing rate of grain size drops, and the pinning effect on the matrix rises to a large extent. When the volume fraction of the primary γ' phase is fixed, the grain growth rate is the lowest when the radius is 1.2μm. The fitting results of grain size, radius and volume fraction of primary γ' show that when the radius of the primary γ' phase is 2.8μm, the minimum fitting $n$ value is 0.23. Simulations in this work also show that the CA model can effectively simulate grain growth in the presence of second phase, and thus can be useful for studying the pinning effect of second phase.